\documentclass[runningheads]{llncs}
\usepackage[T1]{fontenc}
\usepackage{graphicx}
\usepackage{cite}
\usepackage{amsmath,amssymb,amsfonts}
\usepackage{algorithmic}
\usepackage{graphicx}
\usepackage{textcomp}
\usepackage{xcolor}
\usepackage{braket}
\usepackage{listings}
\usepackage{wrapfig}
\usepackage{subfigure}

\newcommand{\F}{\mathcal F}

\def\h{\hat H}

\begin{document}
\title{
Role of Riemannian geometry in double-bracket quantum imaginary-time evolution
}
%
%
\author{
René Zander\inst{1}\orcidID{0000-0003-0603-5637} 
\and
Raphael Seidel\inst{1}\orcidID{0000-0003-3560-9556} 
\and
Li Xiaoyue\inst{2}\orcidID{0009-0004-6879-1598} 
\and
Marek Gluza\inst{2}\orcidID{0000-0003-2836-9523}}
\authorrunning{R. Zander et al.}
%
\institute{
Fraunhofer Institute FOKUS, Berlin, Germany\\
\email{\{rene.zander, raphael.seidel\}@fokus.fraunhofer.de}
\and
School of Physical and Mathematical Sciences, Nanyang Technological University, 21 Nanyang Link, 637371 Singapore, Republic of Singapore\\
\email{marekludwik.gluza@ntu.edu.sg}
}
\maketitle              
\begin{abstract}
Double-bracket quantum imaginary-time evolution (DB-QITE) is a quantum algorithm which coherently implements steps in the Riemannian steepest-descent direction for the energy cost function.
DB-QITE is derived from Brockett's double-bracket flow which exhibits saddle points where gradients vanish.
In this work, we perform numerical simulations of DB-QITE and describe signatures of transitioning through the vicinity of such saddle points. 
We provide an explicit gate count analysis using quantum compilation programmed in Qrisp. 

\keywords{Quantum computing  \and Riemannian gradients \and quantum imaginary-time evolution \and double-bracket quantum algorithms.}
\end{abstract}
\def\H{\hat H}
\section{Introduction}
\label{sec:Introduction}

Applications of quantum computing  to material science~\cite{quantumcentricmaterialscience} or chemistry~\cite{RMP_QC_chemistry} can be achieved by approximating imaginary-time evolution (ITE) which is defined for an input quantum state $\ket{\Psi_0}$, duration $\tau$, and a Hamiltonian $\h$ as 
\begin{align}
	|\Psi(\tau) \rangle = \frac{e^{-\tau \hat H} |\Psi_0\rangle}  {\| {e^{-\tau \hat H}}|\Psi_0\rangle \| }  \ .
 \label{eq:QITE}
\end{align}
For $\tau\rightarrow \infty$ ITE approaches the ground state $\ket{\lambda_0}$ of $\h$, provided that $\langle \Psi_0|\lambda_0\rangle\neq 0$ ~\cite{GellmannLow}. See Ref.~\cite{gluza_DB_QITE_2024} for a bound on the convergence rate in $\tau$.

Recently, Ref.~\cite{gluza_DB_QITE_2024} proposed a quantum algorithm called DB-QITE which is based on the fact that ITE is a double-bracket flow~\cite{gluza_DB_QITE_2024}.
This analytical approach circumvents the limitations of previous approaches stemming from measurement errors ~\cite{motta2020determining,avqite,mcArdle2019} or post-selection~\cite{fragmented_QITE2024}.
Specifically, by differentiating Eq.~\eqref{eq:QITE} and  considering the density matrix $\Psi(\tau)  = \ket{\Psi(\tau)}\bra{\Psi(\tau)}$ we find
\begin{align}\label{eq: DBF DE}
\frac{\partial \Psi(\tau)}{\partial \tau} = \big[ [\Psi(\tau),\h],\Psi(\tau)\big]\ .
\end{align}
This  matrix-valued ordinary differential equation  is an instance of the well-studied Brockett's double-bracket flow (DBF)~\cite{helmke_moore_optimization}. 
It is known that ITE is a gradient-flow~\cite{HacklGeometry} which can be seen as a corollary to the fact that Brockett's DBFs are gradient-flows too~\cite{moore1994numerical,bloch1990steepest,smith1993geometric,dirr2008lie,kurniawan2012controllability,schulte2011optimal,schulte2008gradient}, see~\cite{helmke_moore_optimization} for an in-depth monograph.
In the following, we introduce the essential concepts from Riemannian geometry and identify the scalar quantities that will allow us to assess the influence of geometry on the operation of the DB-QITE quantum algorithm. We will then explicitly evaluate them using numerical simulations.

\section{Role of Riemannian gradients in ITE}

We now discuss how ITE dynamics in Eq.~\eqref{eq: DBF DE} minimizes the energy via gradient descent on a Riemannian manifold.
First, let us make precise the notion of the specific manifold at hand.
Following Ref.~\cite{helmke_moore_optimization}, we  consider the adjoint-unitary manifold $\mathcal M(A) = \{U AU^\dagger\text{ s.t. } U^{-1}=U^\dagger\}$, which is the set of all matrices generated by evolving a Hermitian operator $A$ by a unitary $U$. 
In other words, $\mathcal M(A)$ arises as a mapping from the manifold of unitary matrices $\mathcal U(d) = \{ U\in \mathbb C^{d\times d}\text{ s.t. }U^{-1}=U^\dagger\}$.
Other works on quantum computing have considered $\mathcal U(d)$ the base manifold~\cite{riemannianflowPhysRevA.107.062421}, but similar gradient operators appear and the form of the Riemannian gradient can be carried over to the ITE DBF too.

Next, we define a loss function that maps points in $\mathcal M(A)$ to non-negative numbers.
Specifically, let $B$ be a Hermitian matrix, which we will refer to as the target matrix.
For any $P\in\mathcal M(A)$ of the form $P=U AU^\dagger$ we define the loss function 
\begin{align}
    \mathcal L_{B}(P)=-\frac{1}{2}\|P-B\|_\text{HS}^2\ .
\end{align}
Then the Riemannian gradient evaluated at $P$
is given by~\cite{helmke_moore_optimization,riemannianflowPhysRevA.107.062421}
\begin{align} \label{eq: grad_riemannian}
\text{grad}_P \mathcal{L}_B(P) = -[[P,B],P]\ .
\end{align}
Subsequently, we define the gradient flow on $\mathcal M(A)$ as a smooth curve of points $A(t)$ stretching from $A(0)=A$ at $t=0$ onwards such that $A(t)$ are the unique solution to the gradient flow equation
\begin{align}
    \partial A(t)/\partial t = - \text{grad}_{A(t)}\mathcal L_B(A(t)) = [[A(t), B],A(t)]\ ,
\end{align}
which is the Brockett's DBF equation.

By using unitary invariance of the Hilbert-Schmidt norm we have~\cite{helmke_moore_optimization}
\begin{align}
    \partial_\tau \mathcal L(\tau) = -\| [ A(\tau), B]\|_\text{HS}^2\ .
\end{align}
This links the dynamics of the cost function along the steepest-descent direction to the magnitude of the bracket of the gradient operator.
For ITE, we set $A = \ket{\Psi_0}\bra{\Psi_0}$ and $B=\h$ so $A(\tau)=\ket{\Psi(\tau)}\bra{\Psi(\tau)}$  and the cost function is given by
    $\mathcal{L}_{\h}(\Psi(\tau)) = - \frac12 \|\Psi(\tau) - \h\|_\text{HS}^2$.
By simple algebra, we simplify this to
\begin{align} \label{eq: DBF cost_}
     \mathcal{L}_{\h}(\Psi(\tau))  = E(\tau) - \frac 12(1+ \|\h\|^2_\text{HS})\ .
\end{align}
Only the ITE energy $E(\tau) = \langle \Psi(\tau) | \h | \Psi(\tau) \rangle
$ plays a role and the second term is immaterial to the optimization.
We remark that while individually both ITE and Brockett's DBF have received a lot of attention, to our knowledge this link established in Ref.~\cite{gluza_DB_QITE_2024} is not widely known.

Using the Leibniz rule and Eq.~\eqref{eq: DBF DE} we find that the energy changes as~\cite{gluza_DB_QITE_2024}
\begin{align}
\partial_\tau E(\tau) = -2 V(\tau)\ ,    \label{eq:fluctuationRefrigeration}
\end{align}
where $V(\tau) = \langle \Psi(\tau) | (\h - E(\tau))^2| \Psi(\tau) \rangle
$ is called the energy fluctuation.
This relation implies that higher energy fluctuations in the state lead to a faster energy decrease.
We have $V(0)=0$ for any eigenstate of $\h$, which are the choices of $\ket{\Psi_0}$ where $E(\tau)$ does not decrease. 
Ref.~\cite{helmke_moore_optimization} explains the linear stability analysis of Brockett's DBF and proves that these equilibrium points are unstable \textit{saddle points} unless we have $\ket{\Psi_0} = \ket{\lambda_0}$.

\section{DB-QITE quantum algorithm}

DB-QITE was formulated using approximations of the ITE DBF with unitary operations that can be realized on a quantum computer in the framework of double-bracket quantum algorithms~\cite{double_bracket2024}.
Specifically, using Eq.~\eqref{eq: DBF DE} we have
\begin{align}
\label{eq gradient}
\ket{\Psi(\tau)}=e^{\tau[\ket{\Psi_0}\bra{\Psi_0},\h]}\ket{\Psi_0} +\mathcal O(\tau^2) \ 
\end{align}
and this unitary can be approximated using a group commutator (GC) formula
\begin{align}\label{eqGC}
    G_s(\hat A,\hat B) = e^{i\sqrt s\hat A}e^{i\sqrt s\hat B}    
    e^{-i\sqrt s\hat A}
    e^{-i\sqrt s\hat B}= e^{-s[\hat A,\hat B]} + \mathcal O(s^{3/2})\ .
\end{align}
Ref.~\cite{gluza_DB_QITE_2024} pointed out that the last unitary of the group commutator has a trivial action $e^{-i\sqrt{s}\ket{\Psi_0}\bra{\Psi_0}}\ket {\Psi_0} = e^{i\sqrt s}\ket {\Psi_0}$ which leads to the definition
\begin{align}\label{eq: GCI state main}
    \ket{\omega_{k+1}} = e^{i\sqrt{s_k}\h} e^{i\sqrt{s_k}\omega_k}   
    e^{-i\sqrt{s_k}\h}\ket{\omega_k}\ .
\end{align}
Finally, let $U_k$ denote the circuit to prepare $\ket{\omega_k}$ from a trivial reference state $\ket 0$, i.e., $\ket{\omega_k} := U_k |0\rangle$. We can now use unitarity to simplify $e^{i\sqrt{s}\omega_k}= U_k e^{i\sqrt{s_k}\ket 0\langle0|}    U_k^\dagger$ and obtain the recursive formula for DB-QITE circuit synthesis: 
\begin{align}
    U_{k+1} = e^{i\sqrt{s_k}\h}U_k e^{i\sqrt{s_k}\ket 0\langle0|}    U_k^\dagger
  e^{-i\sqrt{s_k}\h} U_k\ \ .
  \label{DB-QITE Uk}
\end{align}
Here, $U_0$ can be any unitary such that $\ket{\omega_0}:=U_0\ket 0$ yields a valid approximation to the ground state $\ket{\lambda_0}$ of the input Hamiltonian $\h$. 
However, $U_0$ must be chosen carefully as the subsequent steps of DB-QITE continue to leverage on $U_0$ implicitly. 
The interpretation of tangent spaces suggests that a better approximation of Eq.~\eqref{eq gradient} could be advantageous.

Similarly to Ref.~\cite{robbiati2024double}, we consider the higher-order product formula (HOPF) with $\phi=\frac{\sqrt 5 -1}{2}$ 
\begin{align}
    e^{i\phi\sqrt{s}\hat A}e^{i\phi\sqrt{s}\hat B}e^{-i\sqrt{s}\hat A}e^{-i(1+\phi)\sqrt{s}\hat B}e^{i(1-\phi)\sqrt{s}\hat A}e^{i\sqrt{s}\hat B}
    = e^{-s[\hat A, \hat B]} + \mathcal O(s^{2})\ ,
\end{align}
which is more accurate but uses more operations compared to GC.
This leads to a slight generalization of DB-QITE
\begin{align}
    U_{k+1} = e^{i\phi\sqrt{s_k}\h}e^{i\phi\sqrt{s_k}\omega_k}e^{-i\sqrt{s_k}\h}e^{-i(1+\phi)\sqrt{s_k}\omega_k}e^{i(1-\phi)\sqrt{s_k}\h}U_k \ . 
    \label{HOPF U_k}
\end{align}

Ref.~\cite{gluza_DB_QITE_2024} proved that Eq.~\eqref{DB-QITE Uk} leads to a relation similar to Eq.~\eqref{eq:fluctuationRefrigeration} between  the energy $E_k:=\bra{\omega_k}\h\ket{\omega_k}$ and variance $V_k:=\bra{\omega_k}\h^2\ket{\omega_k}-E_k^2$  given by
\begin{align}
     E_{k+1} \le  E_k - 2s_k V_k + \mathcal O (s_k^{2})\ .
    \label{eq fluctuation-refrigeration main Os2} \
\end{align}
This is a direct repercussion of the underlying Riemannian geometry and we will evaluate these quantities in the following.

\subsection{Quantum compiling for DB-QITE}
\label{sec:Compilation}
Compiling the DB-QITE algorithm requires a variety of algorithmic primitives, whose proper coordination poses challenges from a software engineering perspective. 
To effectively facilitate the implementation and modularize the maintenance, debugging, and optimization of each component of the compilation we use the Qrisp programming framework \cite{seidel_2024_qrisp}. 
In particular, Qrisp provides an automated memory management system, which allows several modules to exchange ancillary qubits without intertwining the code, thus facilitating DB-QITE simulations for relatively large systems. 
Another advantage of Qrisp is that the entire DB-QITE implementation is given by the code in Fig.~\ref{lst:qite_impl}.

\begin{wrapfigure}{r}{0.5\textwidth}
\vspace{-2em}
\begin{lstlisting}[language=python, basicstyle = \scriptsize]
from qrisp import *
        
def QITE(qarg,U_0,exp_H,s,k):

    def conjugator(qarg):
        with invert():
            QITE(qarg,U_0,exp_H,s,k-1)

    def reflection(qarg, t):
        with conjugate(conjugator)(qarg):
            mcp(t,qarg,ctrl_state=0)

    if k==0:
        U_0(qarg)
    else:
        s_ = s[k-1]**0.5
        QITE(qarg, U_0, exp_H, s, k-1)
        with conjugate(exp_H)(qarg, s_):
            reflection(qarg, s_)
\end{lstlisting}
\caption{\label{lst:qite_impl} Qrisp implementation of DB-QITE. \texttt{qarg} is the \texttt{QuantumVariable} which is operated upon, \texttt{U\_0} is a state preparation function, \texttt{exp\_H} is a function, which simulates the Hamiltonian in question. \texttt{s} is the array indicating the schedule and \texttt{k} is the recursion depth.}
\vspace{-2em}
\end{wrapfigure}
The \texttt{with} statements enter so-called \textit{QuantumEnvironments}, which represent higher-order quantum functions \cite{seidel_2024_qrisp}.
We will now discuss the compilation primitives natively available in Qrisp.
Specifically, the DB-QITE unitary \eqref{DB-QITE Uk} involves two types of operations and we need to express each using Clifford + T + RZ.
The first type of operation is Hamiltonian evolutions $e^{i\sqrt{s_k}\h}$ that can be implemented via Trotter-Suzuki decomposition~\cite{PhysRevX.TrotterSuzukiError}. The Hamiltonians considered in Sec.~\ref{sec:Experiments} are two-local so each term can be simulated with two CNOT gates, two RZ gates, and some single-qubit Cliffords for change of basis \cite{Yamamoto_2023}. To optimize the depth, we categorize the terms into layers. Terms belonging to the same layer do not act on the same qubit and can therefore be executed in parallel. The classification is achieved by a heuristic graph-coloring algorithm \cite{RLF}.

The second type of operations in Eq.~\eqref{DB-QITE Uk} is reflection gates $ e^{i\sqrt{s_k} \ket0\bra0}$ which are effectively single-qubit phase gates controlled on the remaining qubits.
Qrisp natively provides a compilation of these gates inspired by Ref.~\cite{balauca_2022}: We compute the ``control'' status of the involved qubits into a freshly allocated qubit, which we set to $\ket{1}$ if all control qubits agree with the given control state, in this case $\ket{0}$. Subsequently, we execute a single phase gate on the ancilla qubit before we finally uncompute the ancilla. For the (un)computation of the ancilla, we modify the procedure given in \cite{balauca_2022} to use Gidney's logical AND \cite{Gidney_2018} to compute the intermediate results.

\begin{figure}
\centering
    \includegraphics[scale=0.35]{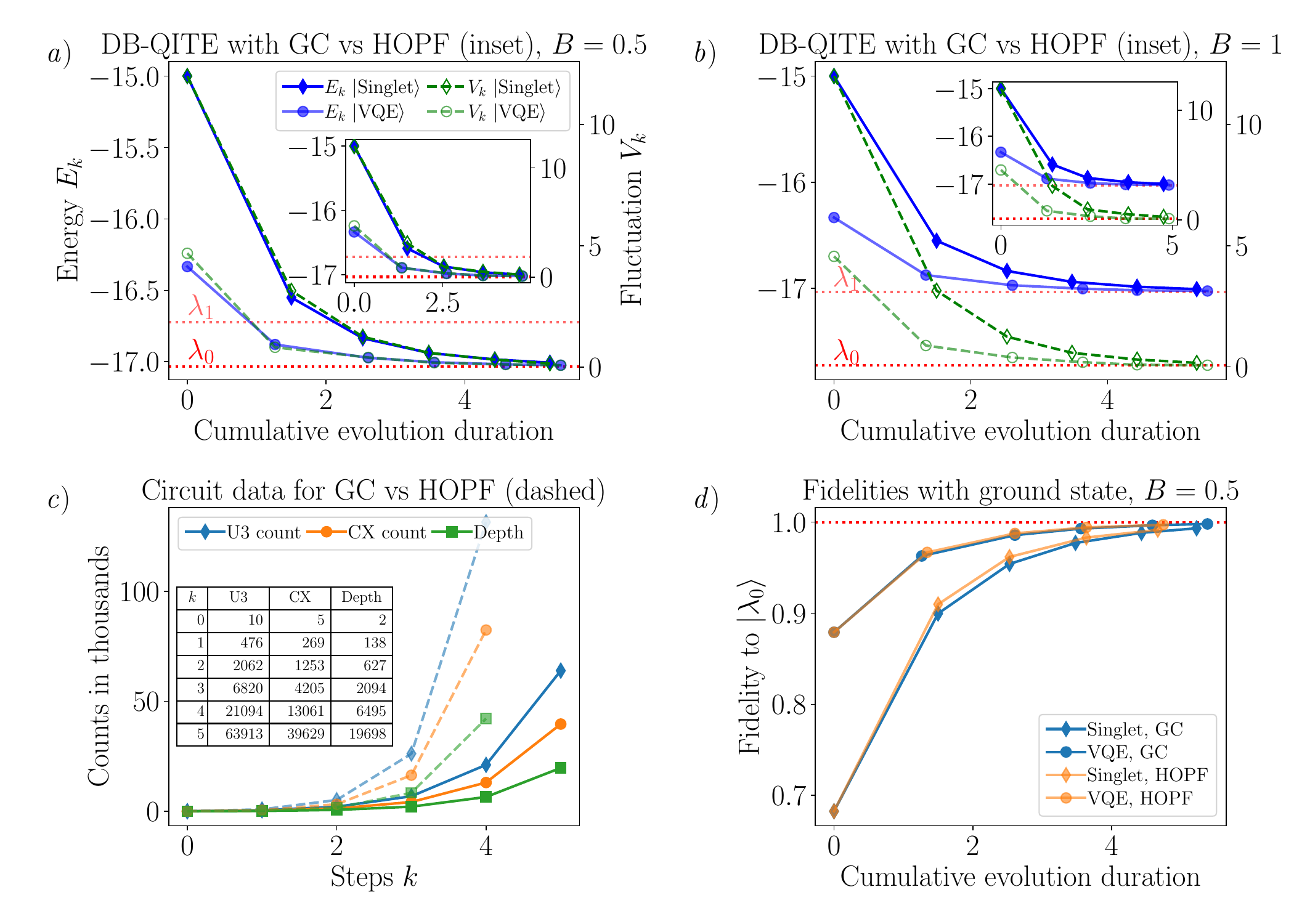} 
    \caption{DB-QITE for the 10-qubit Heisenberg model.
   \textit{ a)} When $B=0.5$ then $\ket{\text{Singlet}}$ has large ground state overlap $\F_0(0)=0.68$ and no overlap with the first excited state $\F_1(0)=0$, and similarly $\ket{\text{VQE}}$ has  $\F_0(0)=0.88$ and $\F_1(0)=0$. DB-QITE rapidly converges to the ground state. 
   \textit{ b)} For $B=1$ ground and first excited states change roles and then $\F_0(0)=0$,  $\F_1(0)=0.68$ for $\ket{\text{Singlet}}$ and $\F_0(0)=0$, $\F_1(0)=0.88$ for $\ket{\text{VQE}}$ and DB-QITE to converges to $\ket{\lambda_1}$. \textit{c)} Counts of  gates $\{\mathrm{U3},\mathrm{CX}\}$ together with the circuit depth for panel a) for $\ket{\text{Singlet}}$ when using GC and HOPF formulas. \textit{d)} GC and HOPF lead to similar fidelity convergence.}
    \label{fig:QITE}
\end{figure}

The overall procedure has a favorable scaling valid in both the short and the long term: The procedure requires\footnote{The classically-controlled Z gate is triggered $50\%$ of times and adds $0.5n$ to runtime.} $3.5n - 4$ entangling gates (which are currently challenging) and only a single phase gate is executed (which is costly in most quantum error-correcting codes). When compiled without mid-circuit measurements, as done in the experiments, the procedure still has a favorable scaling of $6n-6$ entangling gates.
While Ref.~\cite{gluza_DB_QITE_2024} proved that the fidelity to the ground state converges exponentially with the number of DB-QITE steps $k$, Eq.~\eqref{DB-QITE Uk} shows that the circuit depth (i.e., the number of queries to the Hamiltonian simulation, or reflections)  also grows exponentially in $k$. 
Next, we will use Qrisp simulations to clarify this trade-off.

\section{Numerical examples of DB-QITE}
\label{sec:Experiments}

Due to the exponential scaling of the DB-QITE circuit size, only a few recursion steps are practically feasible. To explore the quality of the ground state approximations achievable under this constraint, we implement a fully compilable version of DB-QITE in the Qrisp programming framework, bridging rigorous mathematical theory with application-driven quantum simulations.

Let us consider DB-QITE applied to the transverse field Heisenberg model 
\begin{align}
\label{hamiltonian}
    \h = J\sum\limits_{i=1}^{L}(X_iX_{i+1}+Y_iY_{i+1}+Z_iZ_{i+1})+B\sum\limits_{i=1}^{L}Z_i
\end{align}
where $L$ is the number of qubits and $B$ is interpreted as the magnetic field strength. For the numerical simulations, we set $L=10$, $J=1$, and $B\in\{0.5,1\}$. Hamiltonian simulation, i.e. the unitary $e^{i\tau H}$, is implemented via the second-order Trotter formula with 2 steps. In every DB-QITE step, we use a 20-point grid search to find the $s_k$ that yields the best energy gain.
Additionally, Eq.~\eqref{eqGC} has an approximate invariance $G_{\alpha \beta s}(\hat A /{\alpha}, \hat B /{\beta}) = e^{-s[\hat A, \hat B]} +\mathcal O(s^{3/2})$ but rescaling by $\alpha,\beta$ can influence the approximation constant in $\mathcal O(s^{3/2})$~\cite{double_bracket2024}. We found empirically that setting $\alpha =10, \beta=1$ heuristically allows us to find $s_k$ which yield better ground state approximations.

\subsection{Comparing DB-QITE with GC and HOPF}

Fig.~\ref{fig:QITE} compares the performance of DB-QITE with GC and HOPF in terms of energy drop, energy fluctuations, gate counts, and ground state fidelity. The initial states are constructed by \textit{i)} a tensor product of singlet states $\ket{\text{Singlet}} = 2^{-L/4}(\ket{10}-\ket{01})^{\otimes L/2}$ of consecutive qubits, and \textit{ii)} a VQE warm-start with 1 layer of a problem-specific ansatz \cite{bosse_Heisenberg_2022}. 
The conclusion we can draw is that while HOPF and GC have similar trajectories and can both reach the same level of fidelity (converges to the ground state in \textit{a)} and to the first excited state in \textit{b)}), HOPF requires significantly more gates. 
Therefore, while in general HOPF might facilitate better gradient approximation, it is not required in this case and the simple GC formula suffices.

\subsection{Preconditioned bottlenecks}

\begin{figure}[htbp]
    \centering
    \includegraphics[scale=0.35]{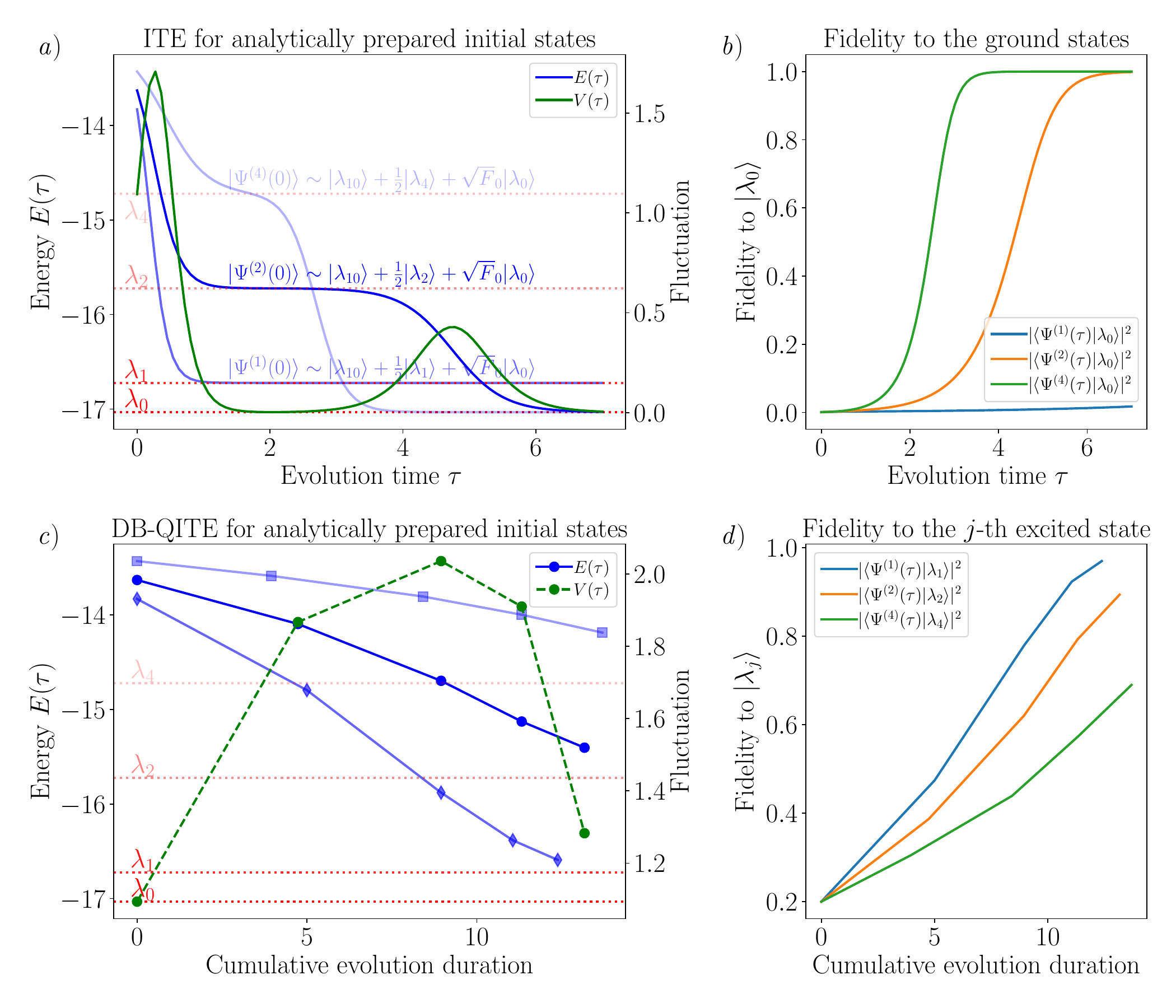} 
    \caption{ ITE \textit{(a,b)} and DB-QITE \textit{(c,d)} for the 10-qubit Heisenberg model with $J=1,B=0.5$ starting from initial states $\ket{\Psi_0^{(j)}}$ with $j=1,2,4$  which are biased to transition from a high-energy state (approximately $\ket{\lambda_{10}}$) through the vicinity of a lower eigenstate $|\lambda_j\rangle$ before reaching the ground state. $|\lambda_j\rangle$ can be interpreted as a saddle point of the ITE because energy decrease can stall almost entirely. 
    \textit{a)} The ITE energy decrease proceeds in three phases. When $\tau$ is small, the energy $E(\tau)$ (shades of blue for $j=1,2,4$) decreases rapidly. As $\tau$ increases, the energy fluctuation $V(\tau)$ (green for $j=2$) drops to $0$ and we reach the saddle point at $\lambda_j$. \textit{b)} The ITE energy becomes stagnant until $\ket{\Psi^{(j)}(\tau)}$ gains about $50\%$ fidelity with the ground state $\ket{\lambda_0}$. However, when the spectral gap is too small such as in the case of $\ket{\Psi^{(1)}(0)}$, leaving the saddle-point vicinity takes far too long time to reach the ground state.  
    \textit{c)} The energy expectations $E_k$ (shades of blue for $j=1,2,4$) with respect to the cumulative QITE duration for different initial states and energy fluctuations $V_k$ for $\ket{\Psi^{(2)}(\tau)}$ (green). Restricting circuit depths to current quantum hardware capacities, we can only explore the first region where the high-energy state phases out and before fully reaching the bottleneck from the underlying saddle point. 
    Energy fluctuations $V_k$ are comparable in magnitude to those in the first peak of ITE.
    \textit{d)} The fidelity to $\ket{\lambda_j}$, i.e. DB-QITE is approaching the bottleneck phase.
    }
    \label{fig:QITE_analytical}
\end{figure}

The upper panels in Fig.~\ref{fig:QITE_analytical} illustrate the role of eigenstates as saddle points of the ITE DBF.
Such obstructions to reaching the ground state $\ket{\lambda_0}$ can appear when the initial state is very close to an eigenstate $\ket{\Psi_0} \approx \ket{\lambda_k}$.
We found, however, that it is possible to transition through the vicinity of an eigenstate $\ket{\lambda_k}$ for initializations that are far from it.
Specifically, let us consider eigenstates of Eq.~\eqref{hamiltonian} and for $k=1,2,4$ set
        $\ket{\Psi_0^{(k)}} = (|\lambda_{10}\rangle + \frac 12 |\lambda_k\rangle + \sqrt{F_0}|\lambda_0\rangle)/\sqrt{1.25 +F_0}$ with $F_0=10^{-6}$.
The $k$'th eigenstate contributes to the initial vector but does not dominate it, i.e. for $\tau=0$ we have $\ket{\Psi_0^{(k)}} \approx 0.9|\lambda_{10}\rangle + 0.45 |\lambda_k\rangle$ and the initial energy $E(0)\approx(\lambda_{10}+\frac{1}{4}\lambda_k)/1.25$.
Under ITE, both components are suppressed exponentially but for the energy scales at hand, we find that after the normalization in Eq.~\eqref{eq:QITE} the contribution of $\ket{\lambda_{10}}$ is almost negligible, i.e. for $\tau \approx 2$ the ITE state $\ket{\Psi^{(k)}(\tau)} \approx  |\lambda_k\rangle $ is almost an eigenstate, as evidenced by the evaluation of the variance.
As $\tau$ increases, this component of the wave functions is eventually suppressed faster than the ground-state component, and we find
$\ket{\Psi^{(k)}(\tau\rightarrow \infty)} \rightarrow  |\lambda_0\rangle $.
However, for $k=1$, we see that leaving the saddle-point vicinity can take an impractically long time.
Note that these plateaus arise from initializations which are \textit{not} exponentially close to a single eigenstate.

The lower panels of Fig.~\ref{fig:QITE_analytical} show DB-QITE results for the same initializations as above, but we remark that in practical quantum computing, preparing such states could be as hard as preparing the ground state itself, if not harder.
That said, we first found the phenomenon exhibited by $\ket{\Psi_0^{(k)}}$ based on a `stuck' VQE initialization which can be prepared by an almost trivial circuit.
As expected, DB-QITE rapidly supresses the high-energy component, but unlike exact ITE, it is unable to reach the true minimum in the realistic regime where the number of steps $k\le 5$. 
This difference can be explained by the fluctuation-refrigeration relation stated in Eq.~\eqref{eq fluctuation-refrigeration main Os2}, which implies that the rate of energy reduction at each step is proportional to the energy variance of the state, which by design is small for $\ket{\Psi_0^{(k)}}$. 
In the discrete setting, low variances in the initial states are much more impeding. 

\section{Discussion and outlook}
\label{sec:Outlook}

Our work demonstrates a systematic method to build quantum circuits for imaginary-time evolution without relying on heuristic variational strategies. 
By leveraging the relationship between DBF and ITE, we explore DB-QITE — a quantum algorithm that iteratively alternates forward and reverse evolutions with reflection operations. 
Our analysis confirms that DB-QITE retains ITE's capability to prepare ground states by minimizing the energy via gradient descent on a Riemannian manifold and the cooling rate is proportional to the energy fluctuations, which have a geometric interpretation as the speed of the gradient flow. 
Additionally, we identify specific cases where DB-QITE may fail to converge to the ground state which could help us understand how to design initializations avoiding bottlenecks.
Using the Qrisp programming framework, we provide numerical examples of DB-QITE with explicit gate counts.

Looking ahead, we plan to explore the performance of DB-QITE on noisy intermediate-scale quantum (NISQ) hardwares and investigate how real-device noise affects the convergence towards accurate ground-state solutions. Another natural extension involves optimizing the circuit designs and parameterization strategies to handle larger system sizes more efficiently. Finally, combining DB-QITE with error-mitigation or classical post-processing techniques could further improve its performance, paving the way for practical quantum simulations of complex many-body systems.

\begin{credits}
\subsubsection{\ackname} MG is supported by the start-up grant of the Nanyang Assistant Professorship at the Nanyang Technological University in Singapore awarded to Nelly Ng.
LX is supported by MOE Tier 1 023816-00001 “Catalyzing quantum security: bridging between theory and practice in quantum communication protocols”.
RS and RZ performed this work in the scope of the PQ-REACT European Union’s Horizon Europe research and innovation program under grant agreement No. 101119547.

\subsubsection{\discintname} The authors have no conflicts of interest to declare that are relevant to the content of this article.
\end{credits}

%
%
%
\bibliographystyle{splncs04}

\begin{thebibliography}{10}
\providecommand{\url}[1]{\texttt{#1}}
\providecommand{\urlprefix}{URL }
\providecommand{\doi}[1]{https://doi.org/#1}

\bibitem{quantumcentricmaterialscience}
Alexeev, Y., Amsler, M., et~al.: Quantum-centric supercomputing for materials science: A perspective on challenges and future directions. Future Generation Computer Systems  \textbf{160},  666--710 (2024).

\bibitem{balauca_2022}
Balauca, S., Arusoaie, A.: Efficient constructions for simulating multi-controlled quantum gates. In: Groen, D., de~Mulatier, C., Paszynski, M., Krzhizhanovskaya, V.V., Dongarra, J.J., Sloot, P.M.A. (eds.) Computational Science -- ICCS 2022. pp. 179--194. Springer International Publishing, Cham (2022).



\bibitem{bloch1990steepest}
Bloch, A.M.: {Steepest descent, linear programming and Hamiltonian flows}. Contemp. Math. AMS  \textbf{114},  77--88 (1990). 

\bibitem{bosse_Heisenberg_2022}
Bosse, J.L., Montanaro, A.: Probing ground state properties of the kagome antiferromagnetic Heisenberg model using the variational quantum eigensolver. Phys. Rev. B  \textbf{105},  094409 (2022). 



\bibitem{PhysRevX.TrotterSuzukiError}
Childs, A.M., Su, Y., et~al.: Theory of Trotter Error with Commutator Scaling. Phys. Rev. X  \textbf{11},  011020 (2021).


\bibitem{dirr2008lie}
Dirr, G., Helmke, U.: Lie theory for quantum control. GAMM-Mitteilungen  \textbf{31}(1),  59--93 (2008). 

\bibitem{GellmannLow}
Gell-Mann, M., Low, F.: Bound states in quantum field theory. Phys. Rev.  \textbf{84},  350--354 (1951). \doi{10.1103/PhysRev.84.350}.

\bibitem{Gidney_2018}
Gidney, C.: Halving the cost of quantum addition. Quantum  \textbf{2}, ~74 (2018).

\bibitem{double_bracket2024}
Gluza, M.: Double-bracket quantum algorithms for diagonalization. Quantum  \textbf{8},  1316 (2024).

\bibitem{gluza_DB_QITE_2024}
Gluza, M., Son, J., et~al.: Double-bracket quantum algorithms for quantum imaginary-time evolution. arXiv preprint arXiv:2412.04554 (2024).

\bibitem{avqite}
Gomes, N., et~al.: Adaptive variational quantum imaginary time evolution approach for ground state preparation. Advanced Quantum Technologies  \textbf{4}(12),  2100114 (2021). 

\bibitem{HacklGeometry}
Hackl, L., Guaita, T., et~al.: Geometry of variational methods: dynamics of closed quantum systems. SciPost Phys.  \textbf{9}, ~048 (2020). 

\bibitem{helmke_moore_optimization}
Helmke, U., Moore, J.B.: Optimization and Dynamical systems. Springer London (2012).


\bibitem{kurniawan2012controllability}
Kurniawan, I., Dirr, G., et~al.: Controllability aspects of quantum dynamics: a unified approach for closed and open systems. IEEE transactions on automatic control  \textbf{57}(8),  1984--1996 (2012).

\bibitem{fragmented_QITE2024}
de~Lima~Silva, T., Taddei, M.M., et~al.: Fragmented imaginary-time evolution for early-stage quantum signal processors. Scientific Reports  \textbf{13},  18258 (2023). 

\bibitem{RLF}
Lotfi, V., Sarin, S.: A graph coloring algorithm for large scale scheduling problems, Computers \& Operations Research \textbf{13}(1),  27--32 (1986).

\bibitem{RMP_QC_chemistry}
McArdle, S., Endo, S., et~al.: Quantum computational chemistry. Rev. Mod. Phys.  \textbf{92},  015003 (2020). 

\bibitem{mcArdle2019}
McArdle, S., et~al.: Variational ansatz-based quantum simulation of imaginary time evolution. npj Quantum Information  \textbf{5}(1), ~75 (2019). 

\bibitem{moore1994numerical}
Moore, J., Mahony, R., et~al.: Numerical gradient algorithms for eigenvalue and singular value calculations. SIAM Journal on Matrix Analysis and Applications  \textbf{15}(3),  881--902 (1994). 

\bibitem{motta2020determining}
Motta, M., Sun, et~al.: Determining eigenstates and thermal states on a quantum computer using quantum imaginary time evolution. Nature Physics  \textbf{16}(2),  205--210 (2020). 

\bibitem{robbiati2024double}
Robbiati, M., Pedicillo, E., et~al.: Double-bracket quantum algorithms for high-fidelity ground state preparation. arXiv preprint arXiv:2408.03987 (2024).


\bibitem{schulte2011optimal}
Schulte-Herbr{\"u}ggen, T., Sp{\"o}rl, A., et~al.: Optimal control for generating quantum gates in open dissipative systems. Journal of Physics B: Atomic, Molecular and Optical Physics  \textbf{44}(15),  154013 (2011).

\bibitem{schulte2008gradient}
Schulte-Herbr{\"u}ggen, T., Glaser, S., et~al.: Gradient flows for optimisation and quantum control: foundations and applications. arXiv preprint arXiv:0802.4195 (2008).

\bibitem{seidel_2024_qrisp}
Seidel, R., Bock, S., et. al: Qrisp: A framework for compilable high-level programming of gate-based quantum computers. arXiv preprint arXiv:2406.14792 (2024).

\bibitem{smith1993geometric}
Smith, S.T.: Geometric optimization methods for adaptive filtering. Harvard University (1993).


\bibitem{riemannianflowPhysRevA.107.062421}
Wiersema, R., Killoran, N.: Optimizing quantum circuits with Riemannian gradient flow. Phys. Rev. A  \textbf{107},  062421 (2023).


\bibitem{Yamamoto_2023}
Yamamoto, T., Ohira, R.: Error suppression by a virtual two-qubit gate. Journal of Applied Physics \textbf{133}(17) (2023).



\end{thebibliography}

%




\end{document}